\documentclass[twocolumn,showpacs,aps,prl]{revtex4}

\usepackage{epsfig}

\begin{document}

\title{Exact Unitary Transformation of the \\
One-Dimensional Periodic Anderson Model}
\author{Raymond Chan and Miklos Gulacsi}
\affiliation{ 
Department of Theoretical Physics,
Institute of Advanced Studies \\
The Australian National University,
Canberra, ACT 0200, Australia}
\date{\today}

\begin{abstract}
An effective hamiltonian is derived exactly for the one-dimensional
periodic Anderson model via a canonical transformation. The canonical
transformation has been calculated up to infinite order, thus an
{\sl exact} transformation was performed in the strict mathematical 
sense. We also discuss briefly the impact of the obtained result on 
understanding the magnetic properties of several Kondo lattice compounds. 
\end{abstract}

\pacs{05.30.Fk, 67.40.Db, 71.10.-w, 71.10.Fd, 71.27.+a}

\maketitle


The periodic Anderson model (PAM) is believed to contain the 
essential physics to describe the low temperature magnetic,
superconducting or semiconducting properties of the heavy
fermion materials and yet the understanding of it is still
very limited. The model is exactly solvable via Bethe Ansatz 
only in the single impurity case \cite{wiegmann} and even in
one-dimension (1D) it's Kondo lattice limit only allows an
exact solution via bosonization \cite{graeme}. 

To begin understanding the role that PAM plays in determining the 
properties of heavy fermion systems we must gain a better understanding 
of the effective interactions and local correlation present in the 
model. We achieved this by applying a canonical transformation to PAM,
similar to that used by Schrieffer and Wolff \cite{SW}, which has
been originally constructed in order to establish the connection
between the single impurity Anderson model and the single impurity 
Kondo model. 

The Schrieffer-Wolff transformation \cite{SW} was performed only up 
to first order, which restricts the validity of the transformation 
to a very limited range of the original parameters. 
In this Letter we focus on the 1D case, where we developed a method
that allows us to derive recursive equations for arbitrary order of
the canonical transformation, and enabled us to sum up the transformation
to infinite order. Thus the canonical transformation is exact, 
and it is not suppose to eliminate any irrelevant degrees of freedom. 
After the transformation is carried out a new Hamiltonian is derived
and the exact microscopic expressions for the effective interaction 
constants obtained. 

From these coupling constants, we will analyse in more detail the 
value of the effective magnetic coupling between the impurities 
(the so-called {\sl Kondo coupling}) to gain a deeper insight into 
the magnetic property of the 1D PAM. This problem has generated
considerable interest since M\"{o}ller and W\"{o}lfe \cite{ferro} 
have shown, using a slave boson approximation, that both ferro- and 
antiferro-magnetism may exists in the 1D PAM. Recent numerical 
calculations \cite{Guerrero} have also shown the existence of 
ferromagnetic phase in the PAM for a wide range of doping ratios 
between quarter-filling and half-filling. On the contrary, 
the ground state of the symmetric 1D PAM is antiferro-magnetic in 
the half-filled case \cite{Ueda} and 
favours antiferro-magnetic interaction in the vicinity of 
quarter filling \cite{ferro}. In addition to this, there are 
many cerium mixed valence materials \cite{batista}, which 
exhibit ferro-magnetism in the presence of strong Kondo-like 
behavior. The exact Kondo coupling, as calculated
hereafter, will shed light on this complex magnetic 
behavior of 1D PAM. 


{\it{1. Canonical Transformation.}} We write the standard PAM Hamiltonian
in 1D as $H = H_{0} + H_{V}$, with
\begin{eqnarray}
H_{0} \: &=& \: - t \sum_{i, \sigma} (c^{\dagger}_{i+1, \sigma}
c^{}_{i, \sigma} +  {\rm h.c.}) 
\: + \: \mu \sum_{i, \sigma} c^{\dagger}_{i, \sigma} c^{}_{i, \sigma}
\nonumber \\
&+& \: \epsilon_{f} \sum_{i, \sigma} f^{\dagger}_{i, \sigma} f^{}_{i, \sigma} 
\: + \: U \sum_{i} n^{f}_{i, \sigma} n^{f}_{i, -\sigma} \; ,
\label{eq:H(0)} 
\end{eqnarray}
and $H_{V} = V \sum_{i \sigma} ( f^{\dagger}_{i \sigma}
c^{}_{i \sigma} + {\rm h.c} )$. 
Here $c^{\dagger}_{i, \sigma }$ ($c^{}_{i, \sigma}$) create
(annihilate) conduction electrons with spin $\sigma$ at lattice 
site $i$, $f^{\dagger}_{i, \sigma}$ ($f^{}_{i, \sigma}$) create 
(annihilate) $f$-orbital impurity electrons ($n^{f}_{i, \sigma} 
= f^{\dagger}_{i, \sigma} f^{}_{i, \sigma}$), and $\mu$ 
($\epsilon_{f}$) are the energies of the conduction ($f$-orbital)
electrons, respectively. $t$ is the nearest-neighbour hopping 
matrix element, $U$ the on-site Coulomb repulsion of the $f$-electrons, 
and $V$ is the strength of the on-site hybridization matrix element 
between electrons in the $f$-orbitals and the conduction band. 

Its unitary transform is simply written as: 
\begin{eqnarray}
\tilde{H} \: &=& \: e^{{\cal S}} (H_{0} + H_{V}) e^{-{\cal S}} 
\nonumber \\
&=& \: H_{0} \: + \: [S,H_{V}] / 2 \: + \: [S,[S,H_{V}]] / 3 
\: + \: \ldots \; , 
\end{eqnarray}
where ${\cal S}$ is determined by the condition 
$H_{V} + [S, H_{0}] = 0$. One can easily verify that 
\begin{equation}
{\cal S} \: = \: \sum_{i \sigma} V ( A + Z f^{\dagger}_{i, -\sigma}
f^{}_{i -\sigma} ) \: ( f^{\dagger}_{i, \sigma} c^{}_{i, \sigma} -
c^{\dagger}_{i, \sigma} f^{}_{i, \sigma}) \; ,
\label{eq:rS}
\end{equation}
with $A = 1 / (2t - \mu + \epsilon_{f})$ and
$Z = 1 / (2t - \mu + \epsilon_{f} + U) - A$ satisfies this
condition. In $H_{0}$ of Eq.\ (\ref{eq:H(0)}) a 
continuum representation (similar to the field theoretical bosonization 
scheme \cite{bosonization}) for the kinetic energy has been used. 

Having found $S$, it is straightforward to determine the first 
($n = 1$), third ($n = 3$) and fifth ($n = 5$) order terms of 
the transformed Hamiltonian which depend on the following commutation,
\begin{equation}
[[S,H_{V}]]_{n} \: = \:
[\overbrace{S,[S,[S,\ldots,[S}^{\mbox{$n$ times}},H_{V}] \ldots] \; ,
\label{eq:nthcomm}
\end{equation}
This commutation was found to have the following form over the first three
odd values of $n$,
\begin{eqnarray}
&& [[S,H_{V}]]_{n} \:
= \: \sum_{i, \sigma} \biggl[
J_{n} (c^{\dagger}_{i, \sigma} c^{}_{i, -\sigma}
f^{\dagger}_{i, -\sigma} f^{}_{i, \sigma}
- n^{c}_{i, \sigma} n^{f}_{i, -\sigma})
\nonumber \\
&& + P_{n} ( c^{\dagger}_{i, \sigma} c^{\dagger}_{i, -\sigma}
f^{}_{i, \sigma} f^{}_{i, -\sigma}  +
f^{\dagger}_{i, \sigma} f^{\dagger}_{i, -\sigma}
c^{}_{i, \sigma} c^{}_{i, -\sigma})
\nonumber \\
&& + G_{n} ( n^{f}_{i, \sigma} - n^{c}_{i, \sigma})
+ I_{n} n^{f}_{i, -\sigma} n^{f}_{i, \sigma}
\nonumber \\
&& + M_{n} n^{f}_{i, -\sigma} n^{c}_{i, -\sigma}
( n^{f}_{i, \sigma} - n^{c}_{i, \sigma})
+ K_{n} n^{c}_{i, -\sigma} n^{c}_{i, \sigma} \biggr] \; ,
\label{eq:nthorder}
\end{eqnarray}

The values of the coefficients $J$, $P$, $G$ and $I$, $K$, $M$ for the 
first three odd $n$ are summarized in Table \ref{tab:1-3OrderResult}. 
Besides the terms which renormalize the starting Hamiltonian, we 
obtained three new effective interactions: $J$, the well-known Kondo 
coupling; $P$, a Josephson type two particle intersite tunnelling; 
$K$, an effective on-site Coulomb repulsion for the conduction 
electrons; and a higher order term, $M$. From these, $K$ and 
$M$ are absent in the first order Schrieffer-Wolff result, as
they depict a higher order coupling between the impurity and
conduction electrons. This coupling becomes stronger as $U$ 
increases, but only to a certain extent before the coupling 
decreases.

It is not by incident that the first three odd order commutation 
Eq.\ (\ref{eq:nthcomm}) is given by a close form Eq.\ (\ref{eq:nthorder}). 
In fact, we will prove in the following using induction that the form of 
the $n$ odd order commutation Eq.\ (\ref{eq:nthcomm}) is given by 
Eq.\ (\ref{eq:nthorder}) in general.

\begin{table}
\caption{\label{tab:1-3OrderResult}The coefficients $J$, $P$, $I$, 
$K$ and $M$ ($a^{\alpha \beta} \equiv A^{\alpha} Z^{\beta}$) of 
the transformed  Hamiltonian in the first, third and fifth order, 
corresponding to the first, second and third row. The $G$ coefficient 
has only one term in each order: 2$A$ in the first, -8$A^3$ in the 
second and 32$A^5$ in the fifth order, respectively.} 
\begin{ruledtabular}
\begin{tabular}{cccccc}
$J$ & $P$ & $I$ & $K$ & $M$ \\
\hline
2$a^{01}$&-$a^{01}$&2$a^{01}$& 0& 0\\
\hline
-32$a^{21}$&-8$a^{21}$&-28$a^{21}$&-4$a^{21}$&16$a^{12}$\\
-32$a^{12}$&-8$a^{12}$&-36$a^{12}$&+4$a^{12}$&+8$a^{03}$\\
-16$a^{03}$&+2$a^{03}$&-16$a^{03}$&          &          \\
\hline
  512$a^{41}$& 224$a^{41}$& 336$a^{41}$& 176$a^{41}$&-352$a^{32}$\\
+1024$a^{32}$&+448$a^{32}$&+848$a^{32}$&+176$a^{32}$&-528$a^{23}$\\
+1024$a^{23}$&+328$a^{23}$&+936$a^{23}$& +88$a^{23}$&-368$a^{14}$\\
 +512$a^{14}$&+104$a^{14}$&+520$a^{14}$&  -8$a^{14}$& -96$a^{05}$\\
 +128$a^{05}$&  -4$a^{05}$&+128$a^{05}$     &      &             \\
\end{tabular}
\end{ruledtabular}
\end{table}

Other than the form of the commutation, one can also find   
a pattern in the coefficients  of the common terms among the
first, third and fifth orders, from Table \ref{tab:1-3OrderResult}. 
This pattern has been verified true up to eleventh order and can 
be proven valid for any order by the same induction. 


{\it{2. Proof by Induction.}} The $n = 1$ case is the well known Schrieffer 
and Wolff result \cite{SW} which represents the first rows of Table
\ref{tab:1-3OrderResult}. By visual inspection, 
it can be observed that there is a pattern in the coefficients  of the 
common terms among the first, third and fifth orders. This pattern can 
be proven to exist for any order by induction, if the same pattern remains 
after commuting Eq.\ (\ref{eq:nthorder}) with $S$ twice. 

To do this, two different indices are introduced to
differentiate the order of the commutation $n$ from the recurrence
of the coefficients $J_{m}$, $P_{m}$, $\ldots$ over odd orders. The
mapping of the two sequences can be written as $n = 2 m + 1$ for
odd order $n$. Assuming that Eq.\ (\ref{eq:nthorder}) is true for
any $m$, we calculate its commutation with $S$ for $[[S,H_{V}]]_{n+1}$,
which yields: 
\begin{widetext}
\begin{eqnarray}
&& \sum_{i, \sigma} \biggl[ - 2 V J_{m} ( A + Z n^{f}_{i, -\sigma} )
\: (n^{f}_{i, -\sigma} - n^{c}_{i, -\sigma} )
- 2 V P_{m} ( A + Z n^{c}_{i, -\sigma} ) \:
(n^{f}_{i, -\sigma} - n^{c}_{i, -\sigma} )
- 2 V G_{m} ( A + Z n^{f}_{i, -\sigma} )
\nonumber \\
&& - 2 V M_{m} ( A + Z ) n^{f}_{i, -\sigma} n^{c}_{i, -\sigma}
- 2 V I_{m} ( A + Z n^{f}_{i, -\sigma}) n^{f}_{i, -\sigma}
+ 2 V K_{m} ( A + Z n^{f}_{i, -\sigma} ) n^{c}_{i, -\sigma}
\biggr] \: \biggl(
c^{\dagger}_{i, \sigma} f^{}_{i, \sigma}
+ f^{\dagger}_{i, \sigma} c^{}_{i, \sigma} \biggr) \; .
\label{eq:n+1thorder}
\end{eqnarray}
\end{widetext}

In the next step of the proof, calculating the commutator of 
Eq.\ (\ref{eq:n+1thorder}) with $S$ again reveals that the obtained
$[[S,H_{V}]]_{n+2}$ has exactly the same form as $[[S,H_{V}]]_{n}$,
with coefficients:
\begin{eqnarray}
J_{m+1} &=& -J_{m} 4V^{2} ( (A + Z)^{2} + A^{2})
- P_{m} 8V^{2} (A^{2} + A Z)
\nonumber \\
&& - G_{m} 4V^{2}(2 A Z + Z^{2})
- I_{m} 4V^{2}(A + Z)^{2} 
\nonumber \\
&& - K_{m} 4V^{2} A^{2} \; ,
\nonumber \\
P_{m+1} &=& - J_{m} 4V^{2} (A^{2} + A Z) - P_{m} 2V^{2}
( (A + Z)^{2} + A^{2})
\nonumber \\
&& - (I_{m} + K_{m}) 2V^{2} (A + Z) A \; ,
\nonumber \\
I_{m+1} &=& -J_{m} 4V^{2} (A + Z)^{2} - P_{m} 4V^{2} (A^{2} + A Z)
\nonumber \\
&& - G_{m} 4V^{2} (2 A Z + Z^{2}) - I_{m} 4V^{2} (A + Z)^{2} \; ,
\nonumber \\
K_{m+1} &=& - J_{m} 4V^{2} A^{2} - P_{m} 4V^{2} (A^{2} + A Z)
- K_{m} 4V^{2} A^{2} \; ,
\nonumber \\
M_{m+1} &=& + K_{m} 4V^{2} (2 A Z + Z^{2}) - M_{m} 4V^{2} (A + Z)^{2}
\nonumber \\
&& + J_{m} 4V^{2} (2 A Z + Z^{2}) \; ,
\nonumber \\
G_{m+1} &=& - G_{m} 4V^{2} A^{2} \; .
\label{eq:Gn+1}
\end{eqnarray} 
This proves that Eq.\ (\ref{eq:nthorder}) is true for any $m$. 
Thus, using mathematical  induction, we conclude that the form 
of the $n$th commutation of  $H_{V}$ with $S$ is closed and
is always given by Eq.\ (\ref{eq:nthorder}). 

{\it{4. Evaluating the Coefficients.}} The recursive equations 
(\ref{eq:Gn+1}) can be solved simultaneously to
give the odd order coefficients of the transformed Hamiltonian,
$J_{m}$, $P_{m}$, $I_{m}$, $K_{m}$ and $M_{m}$.

These recursive equations can be summarized into an matrix form in which
\begin{equation}
\left( \begin{array}{c}
J_{m+1}    \\
2 P_{m+1}  \\
I_{m+1}    \\
K_{m+1}    \end{array} \right)
\: = \: - 4 V^{2} {\bf M} \cdot
\left( \begin{array}{c}
J_{m}  \\
P_{m}  \\
I_{m}  \\
K_{m}  \end{array} \right)
- 4 V^{2}
\left( \begin{array}{c}
\alpha^{2}-\beta^{2} \\
0                    \\
\alpha^{2}-\beta^{2} \\
0                    \end{array} \right)
G_{m} \; , \label{eq:MatrixEq}
\end{equation}
where $\alpha = A + Z$ and $\beta = A$. ${\bf M}$ is a matrix given by:
\begin{equation}
{\bf M} \: = \: \left( \begin{array}{cccc}
\alpha^{2}+\beta^{2} & 2\alpha\beta & \alpha^{2} & \beta^{2} \\
2\alpha\beta & \alpha^{2}+\beta^{2} & \alpha\beta & \alpha\beta \\
\alpha^{2} & \alpha\beta & \alpha^{2} & 0 \\
\beta^{2} & \alpha\beta & 0 & \beta^{2}
\end{array} \right) \; .
\end{equation}

These equations form a set of simultaneous recursive equations with 
the first order, ie, the Schrieffer and Wolff \cite{SW} result as the 
initial condition, and their solution is:
\begin{eqnarray}
J_{m} &=& (-2^{3}V^{2})^{m} (\alpha^{2}+\beta^{2})^{m} 
2 (\alpha-\beta) V^{2} \; ,
\nonumber \\
P_{m} &=& (-2V^{2})^{m} (\alpha^{2}+\beta^{2})^{m-1} 
[2^{2m+1}\alpha\beta 
\nonumber \\
&& - (\alpha+\beta)^{2}] (\alpha-\beta) V^{2} \; ,
\nonumber \\
G_{m} &=& (-2V^{2})^{m}  (2\beta^{2})^{m}  2 \beta V^{2} \; ,
\nonumber \\
K_{m} &=& \{ (-2V^{2})^{m}  (\alpha^{2} + \beta^{2})^{m-1} 
[4^{m}\beta(\alpha-\beta) 
\nonumber \\
&& - \alpha(\alpha+\beta)]  + (-2V^{2})^{m} 
(2\beta^{2})^{m} \}  2 \beta V^{2} \; ,
\nonumber \\
I_{m} &=& (-2^{3}V^{2})^{m}  (\alpha^{2}+\beta^{2})^{m-1} 
2 \alpha^{2} (\alpha-\beta) V^{2}
\nonumber \\
&& + (-2V^{2})^{m}  (\alpha^{2} + \beta^{2})^{m-1} 
2 \alpha \beta (\alpha+\beta) V^{2}
\nonumber \\
&& - (-2V^{2})^{m}  (2\beta^{2})^{m}  2 \beta V^{2} \; ,
\nonumber \\
M_{m} &=& -(-2^{3}V^{2})^{m}  (\alpha^{2}+\beta^{2})^{m-1} 
2 (\alpha-\beta) (\alpha^{2}-\beta^{2}) V^{2}
\nonumber\\
&&- (-2V^{2})^{m}  (\alpha^{2}+\beta^{2})^{m-1} 
4 \alpha \beta (\alpha+\beta) V^{2}
\nonumber \\
&& + (-4\alpha^{2}V^{2})^{m}  2 \alpha V^{2} 
+ (-4\beta^{2}V^{2})^{m}  2 \beta V^{2} \; .
\nonumber
\end{eqnarray}

The even order coefficients can then be deduced from the odd order
coefficients, by applying another commutation. These coefficients 
also have a form, as one would expect, but the pattern is quite 
different to that of odd order. The $(n+1)^{th}$ commutation 
result can be summarized in the form:
\begin{eqnarray}
&& [[S,H_{V}]]_{n+1} = \sum_{i, \sigma}
(R_{m} + S_{m}n^{f}_{i, -\sigma} + T_{m}n^{c}_{i, -\sigma}
\nonumber \\
&& + Q_{m}n^{f}_{i, -\sigma}n^{c}_{i, -\sigma}) \:
( c^{\dagger}_{i, \sigma} f^{}_{i, \sigma}
+ f^{\dagger}_{i, \sigma} c^{}_{i, \sigma}) \; ,
\label{eq:n+1order2}
\end{eqnarray}
where
\begin{eqnarray}
R_{m} &=& -(-4V^{2}\beta^{2})^{m} 4 \beta^{2} V^{3} \; ,
\nonumber \\
S_{m} &=& -[-2^{3}V^{2}(\alpha^{2}+\beta^{2})]^{m} 
8 \alpha (\alpha-\beta) V^{3} 
\nonumber \\
&& -[-2V^{2}(\alpha^{2}+\beta^{2})]^{m} 2 \beta (\alpha+\beta) V^{3} 
\nonumber \\
&& +(-4V^{2}\beta^{2})^{m} 4 \beta^{2} V^{3} \; ,
\nonumber \\
T_{m} &=& [-2^{3}V^{2}(\alpha^{2}+\beta^{2})]^{m} 
8 \beta (\alpha-\beta) V^{3} 
\nonumber \\
&& -[-2V^{2}(\alpha^{2}+\beta^{2})]^{m} 2 \alpha (\alpha+\beta) V^{3} 
\nonumber \\
&& +(-4V^{2}\beta^{2})^{m} 4 \beta^{2} V^{3} \; ,
\nonumber \\
Q_{m} &=& [-2^3V^{2}(\alpha^{2}+\beta^{2})]^{m} 
8 (\alpha-\beta)^{2} V^{3} 
\nonumber \\
&& + [-2V^{2}(\alpha^{2}+\beta^{2})]^{m} 
2 (\alpha+\beta)^{2} V^{3} 
\nonumber \\
&& -(-4V^{2}\alpha^{2})^{m} 4 \alpha^{2} V^{3}
- (-4V^{2}\beta^{2})^{m} 4\beta^{2} V^{3} \; .
\nonumber
\end{eqnarray}
Accordingly, we have obtained the result of the commutation of 
$H_{V}$ with $S$ to any order, and hence have all the information
needed to re-build the Hamiltonian after the transformation. 


{\it{4. The Transformed Hamiltonian.}} Using the general expression 
for the $n$th and the $(n+1)^{th}$ commutation
of $S$ with $H_{V}$, the infinite order transformation
can be calculated. The transformed Hamiltonian
comprises $H_{0}$, the sum of the odd order commutations of $S$
with $H_{V}$ and the sum of the even order commutations:
$\tilde{H} = H_{0} + H_{odd} + H_{even}$, where 
$H_{odd} =\sum_{m=0}^{\infty} [1 / (2m+1)! - 1 / (2m+2)!] 
[[S,H_{V}]]_{2m+1}$ is: 
\begin{eqnarray}
H_{odd} &=& \sum_{i, \sigma} J (c^{\dagger}_{i \sigma}
c^{}_{i, -\sigma}f^{\dagger}_{i, -\sigma} f^{}_{i, \sigma} -
n^{c}_{i, \sigma} n^{f}_{i, -\sigma})
\nonumber \\
&& + P ( c^{\dagger}_{i, \sigma} c^{\dagger}_{i, -\sigma}
f^{}_{i, \sigma} f^{}_{i, -\sigma}  +
f^{\dagger}_{i, \sigma} f^{\dagger}_{i, -\sigma}
c^{}_{i, \sigma} c^{}_{i, -\sigma})
\nonumber \\
&& + G ( n^{f}_{i, \sigma} - n^{c}_{i, \sigma})
+ M n^{f}_{i, -\sigma} n^{c}_{i, -\sigma}
( n^{f}_{i, \sigma} - n^{c}_{i, \sigma})
\nonumber \\
&& + I n^{f}_{i, -\sigma} n^{f}_{i, \sigma}
+ K n^{c}_{i, -\sigma} n^{c}_{i, \sigma} \; .
\label{eq:sumoddH}
\end{eqnarray}
$J$, $P$, $G$, $I$, $K$ and $M$ are the summation of the
corresponding $J_{m}$, $P_{m}$, $G_{m}$, $I_{m}$, $K_{m}$ and $M_{m}$
over infinite number of $m$. If  we define $\alpha = A+Z$, $\beta = A$,
$\theta = \sqrt{2V^{2}(\alpha^{2}+\beta^{2})}$,  
$\theta_{\beta} = 2V\beta$,
$\theta_{\alpha} = 2V\alpha$ and $F(x) = \sin x /x + (\cos x - 1)/x^2$
then the exact values of the coupling constants from
Eq.\ (\ref{eq:sumoddH}) are:
\begin{eqnarray}
J &=&  2 (\alpha - \beta) V^{2} F(2 \theta) \; ,
\nonumber \\
P &=& 2 \alpha \beta (\alpha - \beta) V^{2} F(2 \theta)
- (\alpha - \beta) V^{2}
\frac{(\alpha + \beta)^{2}}{\alpha^{2} + \beta^{2}} F(\theta) \; ,
\nonumber \\
G &=& 2 \beta V^{2} F(\theta_{\beta}) \; ,
\nonumber \\
K &=& 2 \beta^{2} V^{2} \frac{\alpha - \beta}{\alpha^{2} + \beta^{2}}
F(2 \theta) - 2 \alpha \beta V^{2}
\frac{\alpha + \beta}{\alpha^{2} + \beta^{2}} F(\theta) + G \; ,
\nonumber \\
I &=& 2 \alpha^{2} V^{2} \frac{\alpha - \beta}{\alpha^{2} + \beta^{2}}
F(2 \theta) + 2 \alpha \beta V^{2}
\frac{\alpha + \beta}{\alpha^{2} + \beta^{2}} F(\theta) - G \; ,
\nonumber \\
M &=& -2 (\alpha - \beta)
\frac{\alpha^{2} - \beta^{2}}{\alpha^{2} + \beta^{2}} V^{2} F(2 \theta)
- 4 \alpha \beta \frac{\alpha + \beta}{\alpha^{2} + \beta^{2}}
V^{2} F(\theta)
\nonumber\\
&& + 2 \alpha V^{2} F(\theta_{\alpha}) +
2 \beta V^{2} F(\theta_{\beta}) \; .
\nonumber
\end{eqnarray}

Similarly, the even order Hamiltonian $H_{even} = \sum_{m=0}^{\infty}
[1 / (2m+2)! - 1 / (2m+3)!] [[S,H_{V}]]_{2m+2}$ can be evaluated 
by substitution, 
\begin{eqnarray}
H_{even} &=& \sum_{i, \sigma} (R + S n^{f}_{i, -\sigma} +
T n^{c}_{i, -\sigma} + Q n^{f}_{i, -\sigma} n^{c}_{i, -\sigma} ) 
\nonumber \\
&& \times (c^{\dagger}_{i, \sigma} f^{}_{i, \sigma} +
f^{\dagger}_{i, \sigma} c^{}_{i, \sigma}) \; ,
\label{eq:sumeven}
\end{eqnarray}
where $R$, $S$, $T$ and $Q$ are the summation of the corresponding
$R_{m}$, $S_{m}$, $Q_{m}$ and $T_{m}$ over infinite number of $m$.
Using the same notations for $\alpha$, $\beta$, $\theta$, 
$\theta_{\beta}$, $\theta_{\alpha}$
as above, and $F'(x) = \sin x / x^3 -
\cos x / x^2$, we obtain:
\begin{eqnarray}
R &=& -4 \beta^{2} V^{3} F'(\theta_{\beta})  \; ,
\nonumber \\
S &=& -8 (\alpha - \beta) \alpha V^{3} F'(2 \theta)
- 2 (\alpha + \beta) \beta V^{3} F' (\theta)  - R  \; ,
\nonumber \\
T  &=& 8 \beta (\alpha - \beta) V^{3} F' (2 \theta)
- 2 \alpha (\alpha + \beta) V^{3} F' (\theta) - R  \; ,
\nonumber \\
Q  &=& 8 (\alpha - \beta)^{2} V^{3} F' (2 \theta)
+ 2 (\alpha + \beta)^{2} V^{3} F' (\theta) 
\nonumber \\
&& - 4 \beta V^{3} F' (\theta_{\beta}) - 4 \beta V^{3}
F' (\theta_{\alpha}) \; .
\nonumber
\end{eqnarray}

In the symmetric case ($\epsilon_{f} = - U / 2$),
one gets $Z = -2A$, $\theta = \theta_{\beta} = - \theta_{\alpha}
= 2AV$, $\alpha = -A$ and $\beta = A$, both the odd and
even order Hamiltonian coefficients simplify considerably.


{\it{5. Discussions.}} We have shown that an unitary transformation 
of the 1D PAM can be calculated to infinite order. The obtained series 
for every coefficient generated by the transformation are convergent. 
Even though the transformed Hamiltonian appears to be more complicated 
than the original one, it gives the exact expression of the effective 
interactions which are valid for any $U$ and $V$. These represent a
vital source of infromation due to the lack of similar, exactly 
soluble theories of PAM. 

\begin{figure} [htbp]
\includegraphics[width=3in]{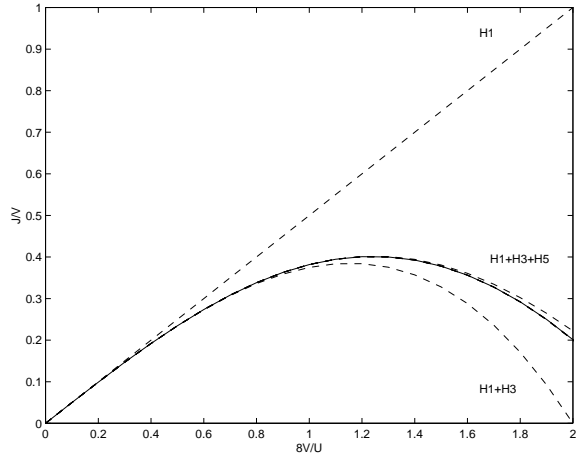}
\caption{\label{fig:result}
The spin coupling constant $J$ up to different order of 
transformation. H1 is the Schrieffer Wolff result, H1+H3 is the result
up to the third order of the transformation, and similarly for H1+H3+H5. 
Solid line shows the result up to infinite order.}
\end{figure}

The Kondo coupling $J$ from our exact result has a behaviour 
different to that
of the well known Schrieffer-Wolff result. This difference is attributed
to the higher order terms of the coupling coefficient $J$ in 
the transformation. As depicted in Fig.\ \ref{fig:result}, these higher
order terms seem to have an effect of suppressing the anti-ferromagnetic 
coupling between the localized electrons for large values of $V/U$.  
As $V/U$ increases, $J$ can change sign, indicating a possible 
change of the magnetic coupling state which has been observed in 
numerical approaches \cite{ferro,Guerrero}. The value of 
$J$ is also crucial to determine \cite{doniach} the Kondo 
($\propto \exp - 1/\vert J \rho \vert)$ and N\'{e}el 
($\propto \vert J \rho \vert^2$) temperatures, where $\rho$
is the density of states for the conduction band at the Fermi level.
In the Kondo regime, see Fig.\ \ref{fig:result}, $J$ passes through 
a maximum, as a function of $U$ and $V$. Accordingly, the Kondo  
and N\'{e}el temperatures will show a similar behavior. 
Such behavior has been observed experimentally  with increasing 
pressure (i.e., $V$) in several cerium Kondo compounds \cite{exp}. 
Hence, this exact result will also shed new light in understanding the 
competition between the Kondo effect and the 
Ruderman-Kittel-Kasuya-Yosida (RKKY) interaction in 1D and revisit 
the Doniach diagram \cite{doniach}. 

Finally, one may notice that the hybridization term is still present in 
the final result of the transformation, see Eq.\ (\ref{eq:sumeven}).
However, the strength of the hybridization is substantially reduced 
to the third order of $V$ as 
$-V((\sin 2VA)/2VA - \cos 2VA) \approx -\frac{1}{3}A^{2}V^{3}$ 
for small $V$. It is in fact possible to remove this hybridization 
term completely, but the transformation involved will be slightly 
different \cite{future}.

\end{document}